\documentclass[preprint,12pt]{elsarticle}

\usepackage{dcolumn}

\usepackage{graphicx}

\biboptions{sort&compress}

\journal{Annals of Physics}

\begin{document}

\begin{frontmatter}

\title{Comment on: ``Solving many-body Schr\"odinger equations with kinetic energy
partition method'', Ann. Phys. 388 (2018) 54-68 by Y-H. Chen and S. D. Chao}

\author{Francisco M. Fern\'{a}ndez}
\ead{framfer@gmail.com}

\address{INIFTA, Divisi\'on Qu\'imica Te\'orica,
Blvd. 113 S/N, Sucursal 4, Casilla de Correo 16, 1900 La Plata, Argentina}

\begin{abstract}
We discuss a recent test of the performance of the kinetic energy partition
method (KEP) through its application to two separable quantum-mechanical
models. We argue that one of the benchmark models is exceedingly simple for
testing any realistic approximate method and that almost any reasonable
approach yields better results. In the second example our exact benchmark
eigenvalues disagree considerably with those chosen by the authors for
comparison, which casts doubts on the accuracy of their KEP approach.
\end{abstract}


\end{frontmatter}

\section{Introduction}

\label{sec:intro}

In the last few years a group of authors have been developing the so called
``split kinetic energy method'' or ``kinetic energy partition method'' (KEP
method from now on)\cite{MC12,MC14,CC17a,CC17b,CC18}. This approach is based
on splitting the kinetic-energy part of a nonrelativistic Hamiltonian in
order to obtain suitable approximations to the energy levels and
corresponding wavefunctions of the Schr\"{o}dinger equation. In general, the
method has been applied to extremely simple nonrealistic models with the
exception of the the hydrogen-molecule ion in the Born-Oppenheimer
approximation\cite{MC12} and the series of two-electron atoms\cite{CC17b}.
The benchmark models are so simple that in some cases\cite{CC17a} they are
even completely unsuitable for the chosen physical applications (even for a
first-order approximation)\cite{F18b}.

In the last paper of the series Chen and Chao\cite{CC18} put forward a
generalized version of KEP that may be suitable for the treatment of
many-body problems, competing, according to the authors, with well stablised
techniques like Hartree-Fock, configuration interaction and density
functional theory, among others. In order to illustrate the supposed
advantages of KEP they apply it to two separable two-body models, one of
which is a trivial textbook example. In this comment we analyze how
realistic are such claims.

In section~\ref{sec:oscillator} we discuss a one-dimensional textbook model
for a two-electron atom in which the attractive nucleus-electron and
repulsive electron-electron interactions are substituted for trivial
harmonic potentials. Section~\ref{sec:harmonium_atom} is devoted to the
so-called Harmonium atom that is a three-dimensional model where only the
nucleus-electron interactions are substituted for harmonic potentials.
Finally, in section~\ref{sec:conclusions} we summarize the main results of
the paper and draw conclusions.

\section{Coupled harmonic oscillators}

\label{sec:oscillator}

The first problem is supposed to represent a one-dimensional atom with
harmonic instead of Coulomb interactions. It is described by the
Schr\"{o}dinger equation $H\psi =E\psi $ with the Hamiltonian operator
\begin{equation}
H=-\frac{\hbar ^{2}}{2m}\left( \frac{\partial ^{2}}{\partial x_{1}^{2}}+%
\frac{\partial ^{2}}{\partial x_{2}^{2}}\right) +\frac{k}{2}\left(
x_{1}^{2}+x_{2}^{2}\right) -\frac{K}{2}\left( x_{1}-x_{2}\right) ^{2},
\label{eq:H_osc_x}
\end{equation}
where $x_{1}$ and $x_{2}$ are the coordinates of the two electrons of mass $%
m $ and $k$ and $K$ are the strengths of the mucleus-electron attraction and
electron-electron repulsion, respectively\cite{CC18}.

In order to simplify the mathematical treatment of any physical problem it
is commonly convenient to resort to dimensionless equations. In the present
case we define the dimensionless coordinates $q_{i}=L^{-1}x_{i}$, $i=1,2$,
where $L=\hbar ^{1/2}/(mk)^{1/4}$, that lead to the dimensionless
Hamiltonian
\begin{equation}
\mathcal{H}=\frac{H}{\hbar \omega }=-\frac{1}{2}\left( \frac{\partial ^{2}}{%
\partial q_{1}^{2}}+\frac{\partial ^{2}}{\partial q_{2}^{2}}\right) +\frac{1%
}{2}\left( q_{1}^{2}+q_{2}^{2}\right) -\frac{\lambda }{2}\left(
q_{1}-q_{2}\right) ^{2},  \label{eq:H_osc_q}
\end{equation}
where, $\omega =\sqrt{k/m}$ and $\lambda =K/k$. Note that $E_{jn}(k,K)=\hbar
\omega \epsilon _{jn}(\lambda )$, where $E_{jn}$ and $\epsilon _{jn}$ are
the eigenvalues of $H$ and $\mathcal{H}$, respectively. The advantage of
using a dimensionless equation is not just that it is simpler than the
original one (the number of model parameters is reduced to a minimum) but
also that we identify the relevant parameters of the system. The common
litany ``we choose $\hbar =m=1$'' will not do the trick. For example, in
this case all the pairs of model parameters $k$ and $K$, treated as
different cases by Chen and Chao\cite{CC18}, are basically the same
mathematical problem if they have the same ratio $K/k$.

This model is a well known textbook example that can be solved exactly. The
reason is that the Schr\"{o}dinger equation is separable by means of the
change of variables
\begin{equation}
\left(
\begin{array}{l}
q_{1} \\
q_{2}
\end{array}
\right) =\mathbf{U}\left(
\begin{array}{l}
Q \\
q
\end{array}
\right) ,\;\mathbf{U}=\frac{1}{\sqrt{2}}\left(
\begin{array}{ll}
1 & 1 \\
1 & -1
\end{array}
\right) ,  \label{eq:change_var_q_osc}
\end{equation}
that leads to a sum of two harmonic oscillators:
\begin{equation}
\mathcal{H}=-\frac{1}{2}\left( \frac{\partial ^{2}}{\partial Q^{2}}+\frac{%
\partial ^{2}}{\partial q^{2}}\right) +\frac{1}{2}Q^{2}+\left( \frac{1}{2}%
-\lambda \right) q^{2}.  \label{eq:H_osc_Qq}
\end{equation}
Note that this Hamiltonian has bound states only when $\lambda <1/2$, and
the corresponding dimensionless eigenvalues are
\begin{equation}
\epsilon _{jn}(\lambda )=j+\frac{1}{2}+\left( n+\frac{1}{2}\right) \sqrt{%
1-2\lambda },\;j,n=0,1,\ldots .  \label{eq:eigenv_osc_dim}
\end{equation}
If $\varphi _{n}^{HO}(k,u)$ is an eigenfunction of the harmonic oscillator $%
H^{HO}=-\frac{1}{2}\frac{d^{2}}{du^{2}}+\frac{k}{2}u^{2}$ then the
eigenfunctions of $\mathcal{H}$ are of the form $\psi _{jn}(Q,q)=\varphi
_{j}^{HO}(1,Q)\varphi _{n}^{HO}\left( \sqrt{1-2\lambda },q\right) $.

As noted above, this model depends essentially on just one parameter $%
\lambda $ and several pairs of model parameters chosen by Chen and Chao\cite
{CC18} are basically the same case; for example: $%
(K,k)=(0.01,0.02),(0.1,0.2),(1,2)$. It is worth noting that they correspond
to $\lambda =1/2$ for which the exact wavefunction is not square integrable;
however, the KEP solutions derived by those authors appear to be square
integrable.

In order to obtain approximate solutions to this exactly solvable problem
the authors rewrite the Hamiltonian operator as $\mathcal{H}=K_{1}+K_{2}$,
where $K_{1}=H_{1}+H_{12}$, $K_{2}=H_{2}+H_{21}$ and
\begin{eqnarray}
H_{i} &=&-\frac{\partial ^{2}}{\partial q_{i}^{2}}+\frac{1}{2}%
q_{i}^{2},\;i=1,2,  \nonumber \\
H_{12} &=&\frac{1}{2}\frac{\partial ^{2}}{\partial q_{1}^{2}}-\frac{\lambda
}{4}\left( q_{1}-q_{2}\right) ^{2},  \nonumber \\
H_{21} &=&\frac{1}{2}\frac{\partial ^{2}}{\partial q_{2}^{2}}-\frac{\lambda
}{4}\left( q_{1}-q_{2}\right) ^{2}.  \label{eq:H_ij_osc}
\end{eqnarray}
Note that these operators are the dimensionless versions (with $\hbar =m=k=1$%
, $K=\lambda $) of those chosen by Chen and Chao\cite{CC18}. The
eigenfunctions of $\mathcal{H}$ are written in terms of the eigenfunctions
of $H_{i}$ and $H_{ij}$:
\begin{eqnarray}
H_{i}\psi _{i}(q_{i}) &=&E_{i}\psi _{i}(q_{i}),\;i=1,2,  \nonumber \\
H_{ij}\psi (\rho ) &=&E_{ij}\psi _{ij}(\rho ),\;i\neq j=1,2,\;\rho
=q_{1}-q_{2}.  \label{eq:psi_ij_osc}
\end{eqnarray}
They first obtain approximate eigenfunctions of $K_{i}$, $i=1,2$ as linear
combinations
\begin{equation}
\phi _{i}(q_{i};q_{j})=C_{i}\psi _{i}(q_{i})+C_{ij}\psi _{ij}(\rho ),\;i\neq
j=1,2.
\end{equation}
Finally, the approximate eigenfunctions are chosen to be
\begin{equation}
\psi (q_{1},q_{2})=\phi _{1}(q_{1};q_{2})\phi _{2}(q_{2};q_{1}).
\label{eq:psi(q1,q2)_approx_osc}
\end{equation}
The authors apply this approach to the ground state and obtain rather
complicated expressions with integrals that involve error functions. In our
opinion a straightforward Hartree approach, followed by configuration
interaction, appears to be simpler and more systematic. However, in what
follows we resort to an even simpler approach.

Note that the authors' approximate wavefunction (\ref
{eq:psi(q1,q2)_approx_osc}) for the ground state depends on $q_{1}^{2}$, $%
q_{2}^{2}$ and $(q_{1}-q_{2})^{2}$. Using the same information (and the
symmetry of the system) we propose the variational function
\begin{equation}
\varphi ^{V}(q_{1},q_{2})=\exp \left[ -\alpha \left(
q_{1}^{2}+q_{2}^{2}\right) -\beta (q_{1}-q_{2})^{2}\right] ,
\label{eq:psi^V_osc}
\end{equation}
where $\alpha $ and $\beta $ are variational parameters. The variational
integral is
\begin{equation}
W(\alpha ,\beta )=\frac{\left\langle \varphi ^{V}\right| \mathcal{H}\left|
\varphi ^{V}\right\rangle }{\left\langle \varphi ^{V}\right. \left| \varphi
^{V}\right\rangle }=\frac{4\alpha ^{3}+12\alpha ^{2}\beta +\alpha \left(
8\beta ^{2}-\lambda +1\right) +\beta }{4\alpha \left( \alpha +2\beta \right)
},  \label{eq:W_var_osc}
\end{equation}
and the variational conditions $\partial W/\partial \alpha =0$ and $\partial
W/\partial \beta =0$ lead to
\begin{eqnarray}
4\alpha ^{4}+16\alpha ^{3}\beta +\alpha ^{2}\left( 16\beta ^{2}+\lambda
-1\right) -2\alpha \beta -2\beta ^{2} &=&0,  \nonumber \\
4\alpha ^{2}+16\alpha \beta +16\beta ^{2}+2\lambda -1 &=&0.
\label{eq:var_cond_osc}
\end{eqnarray}
The optimal solution to the latter system of equations is
\begin{equation}
\alpha _{opt}=\frac{1}{2},\;\beta _{opt}=\frac{\sqrt{1-2\lambda }}{4}-\frac{1%
}{4},
\end{equation}
that leads to the exact ground-state energy
\begin{equation}
W(\alpha _{opt},\beta _{opt})=\epsilon _{00}(\lambda )=\frac{1}{2}+\frac{%
\sqrt{1-2\lambda }}{2}.
\end{equation}
Note that we have not resorted to the knowledge of the exact solution in
order to build the variational function (\ref{eq:psi^V_osc}) but to the form
of the potential given in (\ref{eq:H_osc_q}) or to the form of the KEP
function (\ref{eq:psi(q1,q2)_approx_osc}). The problem with this trivial
example is that most reasonable trial functions will lead to the exact
result. In our opinion the KEP approach leads to the most complicated ones
and it is unclear how to improve them to obtain better results.

\section{Harmonium atom}

\label{sec:harmonium_atom}

The second example is somewhat more challenging as it represents two
electrons in a space of three dimensions with the Hamiltonian
\begin{equation}
H=-\frac{\hbar ^{2}}{2m}\left( \nabla _{1}^{2}+\nabla _{2}^{2}\right) +\frac{%
k}{2}\left( r_{1}^{2}+r_{2}^{2}\right) +\frac{e^{2}}{r_{12}},
\label{eq:H_harmonium}
\end{equation}
where $m$ and $e$ are the mass and charge of an electron and $k$ is a
suitable force constant for the harmonic nucleus-electron attractive part of
the potential. As in the preceding example we choose dimensionless
coordinates $\mathbf{q}_{i}=L^{-1}\mathbf{r}_{i}$, also with $L=\hbar
^{1/2}/(mk)^{1/4}$, and the resulting dimensionless Hamiltonian is
\begin{equation}
\mathcal{H}=\frac{H}{\hbar \omega }=-\frac{1}{2}\left( \nabla
_{q_{1}}^{2}+\nabla _{q_{2}}^{2}\right) +\frac{1}{2}\left(
q_{1}^{2}+q_{2}^{2}\right) +\frac{\lambda }{q},
\label{eq:H_harmonium_dimensionless}
\end{equation}
where $q=|\mathbf{q}_{1}-\mathbf{q}_{2}|=L^{-1}\left| \mathbf{r}_{1}-\mathbf{%
r}_{2}\right| $, $\omega =\sqrt{k/m}$ and $\lambda =m^{3/4}e^{2}\hbar
^{-3/2}k^{-1/4}$.

The Schr\"{o}dinger equation is separable in terms of the variables
\begin{equation}
\mathbf{Q}=\frac{1}{2}\left( \mathbf{q}_{1}+\mathbf{q}_{2}\right) ,\;\mathbf{%
q}=\mathbf{q}_{1}-\mathbf{q}_{2},
\end{equation}
that lead to the new Hamiltonian
\begin{eqnarray}
\mathcal{H} &=&\mathcal{H}_{Q}+\mathcal{H}_{q},  \nonumber \\
\mathcal{H}_{Q} &=&-\frac{1}{4}\nabla _{Q}^{2}+Q^{2},\;\mathcal{H}%
_{q}=-\nabla _{q}^{2}+\frac{1}{4}q^{2}+\frac{\lambda }{q}.
\end{eqnarray}
The eigenfunctions are of the form $\psi
_{\{j\},\{n\}}=f_{\{j\}}(Q)g_{\{n\}}(q)$, where $\{j\}$ and $\{n\}$ are two
sets of suitable quatum numbers and $f_{\{j\}}(Q)$ and $g_{\{n\}}(q)$ are
eigenfunctions of $\mathcal{H}_{Q}$ and $\mathcal{H}_{q}$, respectively. The
Schr\"{o}dinger equation for the three-dimensional isotropic harmonic
oscillator $\mathcal{H}_{Q}$ is exactly solvable and that for $\mathcal{H}%
_{q}$ should be solved approximately, except for some particular values of $%
\lambda $ for which there are exact analytical results (see, for example,
reference~\cite{CP00} and the bibliography therein).

In particular, the ground-state eigenvalue is given by
\begin{equation}
E_{0}=\hbar \omega \left[ \frac{3}{2}+\epsilon _{0}(\lambda )\right] ,
\label{eq:E_0_atom}
\end{equation}
where $\epsilon _{0}(\lambda )$ is the lowest eigenvalue of $\mathcal{H}_{q}$%
. As noted above, the advantage of using a dimensionless equation is that we
do not have to bother with most of the physical constants. However, in order
to compare present results with those of Chen and Chao\cite{CC18} we set $%
\hbar =m=e=1$ so that equation (\ref{eq:E_0_atom}) becomes
\begin{equation}
E_{0}(k)=\sqrt{k}\left[ \frac{3}{2}+\epsilon _{0}\left( k^{-1/4}\right)
\right] .  \label{eq:E_0(k)_atom}
\end{equation}

\begin{figure}[tbp]
\begin{center}
\includegraphics[width=9cm]{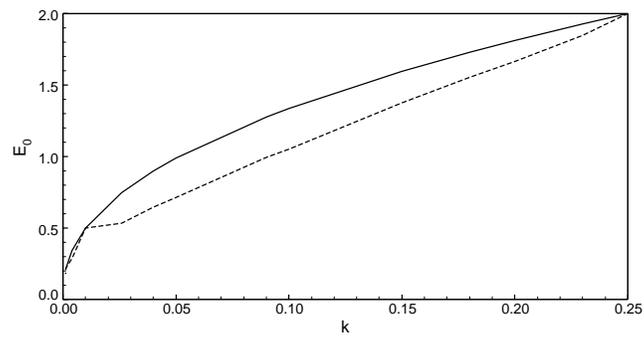}
\end{center}
\caption{Present ground-state eigenvalues for the harmonium atom (solid
line) and those of Chen and Chao\protect\cite{CC18} (dashed line)}
\label{fig:KEPAT}
\end{figure}

We have calculated the eigenvalue $\epsilon _{0}(\lambda )$ by means of the
Riccati-Pad\'{e} method\cite{FMT89a, FMT89b} that converges so fast that
yields extremely accurate results\cite{FG17}. In order to test them we also
tried the Rayleigh-Ritz variational method\cite{P68} with the unnormalized
basis set $f_{j}=q^{j}\exp \left( -\frac{q^{2}}{4}\right) $, $j=0,1,\ldots $
(suitable for $s$ states). This approach is known to yield upper bounds to
all the eigenvalues\cite{P68} (and references therein). Both sets of results
agree perfectly and in Table~\ref{tab:HA} we show the eigenvalue $E_{0}(k)$
for those values of $k$ chosen by Chen and Chao\cite{CC18}. Present results
an theirs disagree considerably except for the exactly solvable cases $k=0.25
$ ($\lambda =\sqrt{2}$) and $k=0.01$ ($\lambda =\sqrt{10}$). The two
approaches used here yield the exact analytical results in such particular
cases: $\epsilon _{0}=5/2$ and $\epsilon _{0}=7/2$, respectively. Figure~\ref
{fig:KEPAT} shows our smooth results and those of Chen and Chao\cite{CC18}
that exhibit a suspicious jagged behaviour. For this reason and for the fact
that our results were obtained by means of two completely different
approaches that agree to the last digit we are confident of them. On the
other hand, when refering to the ``exact'' results, Chen and Chao merely
point to a paper by other authors\cite{CP00} (which exhibits just the cases $%
k=0.25$ and $k=0.01$ used for comparison) and do not indicate how the
calculation was done. If, as we deem, their benchmark results are
inaccurate, then it appears most \textit{fortunate} that they agree so
closely with the KEP ones. Using present exact results as benchmark the
errors of the KEP eigenvalues rise considerably for most values of $k$.

\section{Conclusions}

\label{sec:conclusions}

The authors promise that they will present an approximate method for the
treatment of many-body quantum-mechanical systems that will rival with
existing approaches and in the end they merely treat two separable models,
one of them exceedingly simple. At first sight this paper appears to be a
step backward in the application of KEP to realistic physics problems if one
compares it with that earlier work on the ground states of helium-like atoms%
\cite{CC17b}. There is no doubt that the first model is no suitable
benchmark for a realistic approximate method. In the second one the authors
apparently failed (that is, of course, our opinion) to obtain accurate
results for comparison and one can certainly doubt about the accuracy of
their KEP method. In fact, if one takes into account present exact results
the KEP errors are considerably larger for some values of $k$.

\begin{table}[tbp]
\caption{Eigenvalues of the harmonium atom calculated by means of the
Riccati-Pad\'e method}
\label{tab:HA}
\begin{center}
\begin{tabular}{ll}
\hline
\multicolumn{1}{c}{$k$} & \multicolumn{1}{c}{$E_0$} \\ \hline
0.2500 & 2 \\
0.2300 & 1.9273546297410884205 \\
0.2000 & 1.8116899843671347580 \\
0.1800 & 1.7292911575097244563 \\
0.1500 & 1.5958054174355393322 \\
0.1000 & 1.3360503187251752778 \\
0.0900 & 1.2760601721269501937 \\
0.0500 & 0.98925143507101418781 \\
0.0400 & 0.89879859060929913976 \\
0.0260 & 0.74778853894031961765 \\
0.0100 & 0.5 \\
0.0040 & 0.34224945694769201625 \\
0.0013 & 0.21689817637450858280 \\
0.0012 & 0.21004123606565067787
\end{tabular}
\end{center}
\end{table}

\end{document}